\documentclass{JHEP3}
\pdfoutput=1
\usepackage[pdftex]{graphics}
\usepackage{amssymb,amsfonts,amsmath,amsthm,latexsym,verbatim,epsfig}
\newcommand{\nn}{\nonumber}
\newcommand{\beq}{\begin{equation}}
\newcommand{\eeq}{\end{equation}}
\newcommand{\bea}{\begin{eqnarray}}
\newcommand{\eea}{\end{eqnarray}}

\newcommand{\1}{ \,  \raisebox{+0.14em}{{\hbox{{\rm \scriptsize ]}} \raisebox{-0.2em}{\kern-.8em\hbox{1}}}} \, }  

\title{Eluding   SUSY  at every  genus on stable closed string vacua.}

\author{Sergio L. Cacciatori$^{a}$ and
Matteo A. Cardella$^{b}$ \\
$^a$ Department of Physics and Mathematics, \\
\hspace*{0.15cm} Universit\`a dell'Insubria, Via Valleggio 11, I-22100 Como, Italy, and \\
\hspace*{0.15cm} INFN, Sezione di Milano, Via Celoria 16, I-20133 Milano, Italy \\
\hspace*{0.15cm} E-mail address: sergio.cacciatori@uninsubria.it.\\
$^b$ Department of Physics, Universit\`a di Milano Bicocca, \\
\hspace*{0.15cm} Piazza della Scienza 3, I-20126 Milano, Italy, \\
\hspace*{0.15cm} E-mail address: matteo@phys.huji.ac.il}

\abstract{In closed string vacua, ergodicity   of unipotent flows provide
a key for relating   vacuum stability  to the UV behavior of  spectra and interactions.
 Infrared finiteness  \emph{at all genera} in perturbation theory can be rephrased  in terms of  cancelations
involving \emph{only tree-level} closed strings scattering amplitudes. This provides quantitative  results  on
 the allowed deviations from supersymmetry    on  perturbative  stable vacua.
 From a mathematical perspective,  diagrammatic relations involving closed  string amplitudes suggest
a relevance of unipotent flows  dynamics  for the Schottky problem and for the construction of the superstring measure.}


\begin{document}

\section{Introduction}

In recent years, mathematical results in  homogenous space dynamics
have been  leading  to striking results in number theory, (see for example \cite{EW},\cite{EMV},\cite{ELMV}).
 In a series of papers \cite{C1},\cite{C2},\cite{CC1},\cite{CC2},\cite{ACER}, we have proved that
 fruitful  interactions   arise also   between  homogenous space dynamics  and string theory.
This research lines lead  to results both in  string theory
\cite{C1},\cite{CC1},\cite{ACER}, and in the  mathematics of the automorphic forms and of unipotent flows in homogenous spaces \cite{C2},\cite{CC2}.

\vspace{.27 cm}

In a well known paper, Kutasov and Seiberg \cite{KS} have shown that in backgrounds with no  tachyons,
closed string spectra  exhibit a global UV asymptotic   Fermi-Bose degeneracy.
They dubbed  this global cancelation among  bosonic and fermionic degrees of  freedom
 as \emph{Asymptotic Supersymmetry}. This UV property
of closed string spectra  is related by modularity to infrared finiteness of
the one-loop amplitude.
At genus one  level, it indicates allowed  deviations from supersymmetry  on a  stable vacuum.
  It was remarked in \cite{C1} that   this UV  property of (one-loop) stable closed string vacua is related to
mathematical theorems on
uniform distribution of long horocycles in the modular surface   $\pmb{SL}(2, \mathbb{Z})  \backslash \pmb{SL}(2,\mathbb{R})$.

\vspace{.27 cm}

It is the purpose of this paper to extend the original one-loop (genus one) analysis of \cite{KS}
to all genera in closed string perturbation theory.
We achieve this goal by using certain  mathematical theorems on the dynamics of multidimensional unipotent flows,
we have recently obtained in \cite{CC2}.
 Once applied to  genus $g$ closed  string  amplitudes,
those results translate into relatively  simple operations at a diagrammatic level.
 They correspond to suitable  cuts  of a Riemann surface handle(s).
 This allows to rewrite  a genus $g$ vacuum amplitude with $g$ handles as
 a sum of genus  $(g - 1)$ two-point functions. Uniformization results in \cite{CC2}
 instruct to  sum over  all  physical string states flowing  through the $(g - 1)$ amplitude two-external legs,
 (figure \ref{SFig6}).
  The sum over physical states is regulated by an ultraviolet cutoff $\Lambda$, and
uniform distribution theorem, (Theorem $2$  in \cite{CC2}), ensures that
 the original vacuum amplitude is recovered in the $\Lambda \rightarrow \infty$ limit.
Moreover, for large $\Lambda$  one has that the error term is
under control \cite{CC2}. Interestingly, the error estimate  is intimately  related to the non trivial zeros of the Riemann zeta function,
and a result for this quantity would prove or disprove the Riemann hypothesis \cite{CC1},\cite{CC2}.

\vspace{.27 cm}

In \cite{CC2} was proved, (Theorem $1$), that  the unipotent average of the string integrand automorphic
function is a modular invariant function under the genus $(g - 1)$
modular group. This is a crucial property, in order to be able to
apply iteratively the uniform distribution theorem.   In a diagrammatic
language  this corresponds
to  cutting  handles in a closed string vacuum  amplitude, and transmute each handle into a pair of external
 legs.
In this way,  one is able to reduce
a genus $g$ vacuum amplitude into a sum over  tree-level amplitudes with
$2g$ external legs, (see again figure \ref{SFig6}).
Infrared finiteness of the genus $g$ vacuum amplitude
is then translated in  constraints involving
 ultraviolet   cancelations among $2g$-point \emph{tree level} amplitudes.
This is the way closed strings can elude SUSY on stable vacua.
In a sense our results provide the completion \emph{at all genera} of the   condition
of \emph{Asymptotic Supersymmetry}, obtained at one-loop level  in \cite{KS}, (see also \cite{Di} for related work).

\vspace{.27 cm}

\section{Genus one: uniform distribution of long horocycles viz cutting the  torus handle}\label{genus1}

\vspace{.27 cm}

The one-loop torus amplitude is given by the following modular integral
\vspace{.27 cm}

\beq
A_1  =  \int_{\mathcal{D}_{1}} dw dv \, v^{-\frac{d}{2} - 1} Str\left( e^{2\pi i w (L_{0} - \bar{L}_{0})} e^{- \pi v (L_{0} + \bar{L}_{0})} \right), \label{A1}
\eeq

\vspace{.27 cm}

\noindent where $\tau = w + iv$, is the worldsheet torus modulus,  $w \in \mathbb{R}$, $v > 0$.
As we shall see, this  notation
for the real and imaginary parts  of $\tau$  reflects Iwasawa coordinatization of
the genus $g = 1$ upper complex plane $\mathcal{H}_1$. In (\ref{A1})
$d$ is the number of non compact space-time directions, the supertrace assigns a minus sign to fermionic closed string states,
$L_0$ and $\bar{L}_0$ are the zero modes of the Virasoro operators, and the integral is performed on
a modular domain  $\mathcal{D}_{1} \sim SL(2,\mathbb{Z}) \backslash \mathcal{H}_{1}$, with $\mathcal{H}_1$ the upper complex plane.

\vspace{.27 cm}

\begin{figure} 
\begin{center}
\includegraphics[width = 8cm]{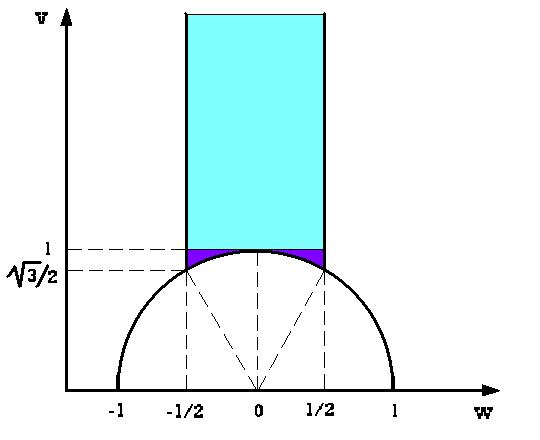}
\caption{The torus modular surface in the upper complex plane.  On the shaded integration  region below $v = 1$,
non-physical closed string states do contribute to the one-loop vacuum energy.}\label{SFig0}
\end{center}
\end{figure}

On the subregion of $\mathcal{D}_1$, where  $v > 1$, (see figure \ref{SFig0}), the  $w$ coordinate  is integrated mod(1).
 This integration
enforces the physical condition of level matching $(L_{0}  -  \bar{L}_{0})| \Phi \rangle = 0$,
 which selects  closed string \emph{physical} states.
The one-loop torus vacuum amplitude $A_1$ in the representation given in (\ref{A1}) receives contributions from \emph{non physical} closed string states, from the integration subregion where  $\sqrt{3}/2 < v < 1$, (the shaded region in figure \ref{SFig0}).
 Since the integration region $\mathcal{D}_1$ does not touch the $\mathcal{H}_1$ boundary $v \rightarrow 0$,
the UV physical  region is not probed, and the vacuum amplitude is free from ultraviolet problems.

\vspace{.27 cm}

\begin{figure}[htbp]
\centering
\includegraphics[height = 12cm, width =  2cm ]{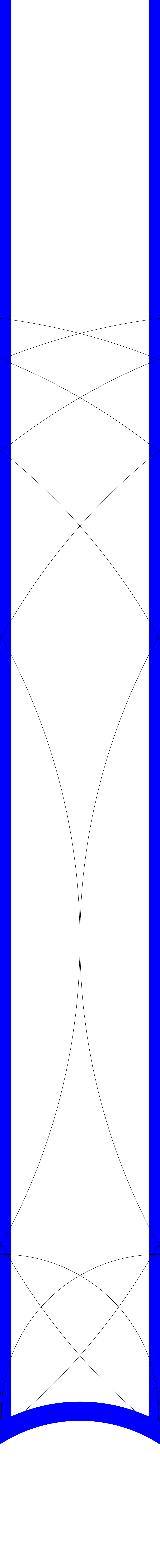}
\hspace{.8 cm}
\includegraphics[height = 12cm, width =  2cm ]{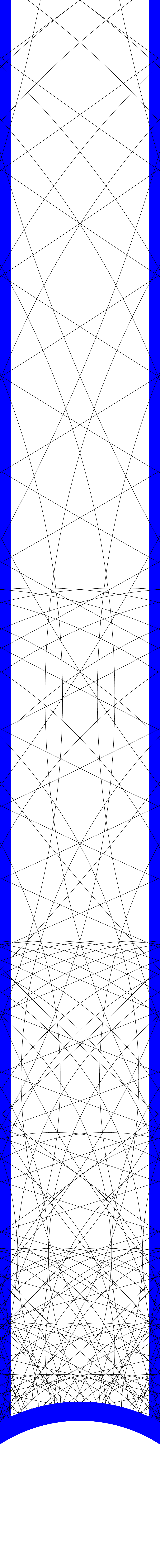}
\hspace{.8 cm}
\includegraphics[height = 12cm, width =  2cm ]{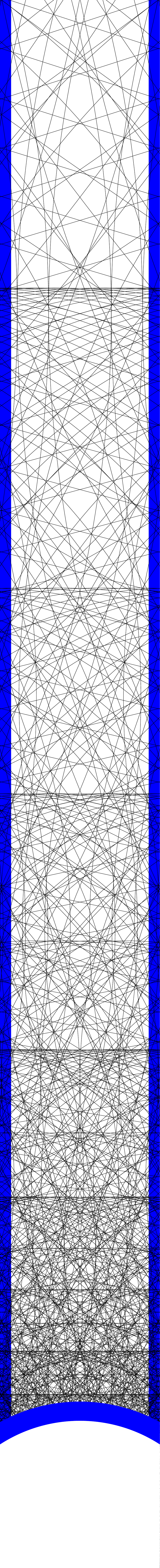}
\caption{ \textbf{Modular images  of  horocycles of increasing length.}
In  the upper complex plane the horizontal line $H_{\alpha} : = \mathbb{R} + i\alpha$
is called a horocycle, since it can be thought as a circle tangent to infinity.
$H_{\alpha}$ is modularly equivalent to a (infinite countable) family of circles, all tangent to
the real axis in rational points. It is interesting to map  $H_{\alpha}$
 in the standard fundamental domain and observe the behavior of its modular image  in the  $\alpha \rightarrow 0$ limit.
 Due to the subgroup $\Gamma_{\infty}$ of the modular group $SL(2,\mathbb{Z})$ given by integral translations
 along the real axis, it is enough to map in the modular domain  the segment  $\Gamma_{\infty} \backslash H_{\alpha} =  [-1/2, 1/2 ) + i\alpha $,
 with hyperbolic length $1/\alpha$.  We refer to this latter quantity  as the length of the horocycle $H_{\alpha}$.
 What happens to the modular image in the standard modular domain  of the horocycle in the increasing length limit $\alpha \rightarrow 0$
 can be observed in figure.
Left:  modular image of the  line $y = \frac{1}{8}$.
Center: modular image of the line  $y = \frac{1}{100}$. Right: modular image
of the line  $y = \frac{1}{400}$.
In all cases the modular domain is truncated up to
a $y \le 10$.   The modular image of a line $y = \alpha$ tends to become dense
in the modular domain as the horizontal line gets close  to the real axis, ($\alpha \rightarrow 0$ limit).
Indeed, in the $\alpha \rightarrow 0$ limit the modular image of the horocycle
 $y = \alpha$ tends to uniformly cover the modular domain  \cite{He}.}
\label{Sflow}
\end{figure}

However, there is an alternative  representation for the genus one vacuum amplitude $A_1$, which allows to probe
the ultraviolet properties of the closed string spectrum. In this representation,  one can check that $A_1$
receives contribution \emph{only} from physical states. This alternative description \cite{C1} follows  by using
ergodic properties of the horocycle flow \cite{He},\cite{Fu},\cite{DS}, (figure \ref{Sflow}). Ergodicity of the horocycle flow states that the modular image of the horocycle  $H_{\alpha} = \mathbb{R} + i\alpha \subset \mathcal{H}_1$, in the  $\alpha \rightarrow 0$ limit  covers uniformly
the modular domain $\mathcal{D}_1$. This implies that for a  continuous bounded  modular function $f = f(w,v)$
its horocycle average $\langle f \rangle_{H_{\alpha}}$ for $\alpha \rightarrow 0$ tends to its average on the modular region $\langle f \rangle_{\mathcal{D}_1}$

\vspace{.27 cm}

\beq
\lim_{\alpha \rightarrow 0} \langle f \rangle_{H_{\alpha}} =    \lim_{v \rightarrow 0} \int_{0}^{1} dw f(w,v) = \frac{1}{Vol(\mathcal{D}_1)} \int_{\mathcal{D}_{1}} dw dv \, v^{-2} \, f(w,v) =  \langle f \rangle_{\mathcal{D}_1}, \label{u}
\eeq

\vspace{.27 cm}

where $Vol(\mathcal{D}_1) = \pi /3$. Notice, that the l.h.s. is indeed the $f$ average  along $H_v$ computed with
the $\mathcal{H}_1$ hyperbolic metric $ds^2 = v^{-2}(dw^2 + dv^2 )$ \footnote{Equation (\ref{u}) holds for every continuous bounded modular function $f$,
however, in string theory,  in the absence of tachyons, generically one has to deal
with modular  functions of polynomial growth at infinity (type II theories), or
of exponential growth at infinity (Heterotic theories). In this latter case,
the modular invariant integrand  function  contains  terms of exponential growth
for $v \rightarrow \infty$, which are removed by $w$ integration mod (1).
 This terms are  dubbed as \emph{unphysical tachyons} \cite{KS}, since they
correspond to tachyonic  states in the supertrace  (\ref{A1}) that do not respect level matching.
In the type II case, eq. (\ref{u}) is proved to hold  \cite{Za2},\cite{ACER},\cite{C2}, while in the heterotic case  eq. (\ref{u}) is expected
to hold \cite{ACER},\cite{C2} on physical grounds, although this has not been actually  proved, (see \cite{C2} for a discussion and
some related  mathematical results on this problem)}.

\vspace{.27 cm}

By applying the uniform distribution result (\ref{u}) to the genus one torus amplitude $A_1$ (\ref{A1}),
one finds

\vspace{.27 cm}

\beq
A_1 =   Vol(\mathcal{D}_1) \lim_{v \rightarrow 0} v^{1 - d/2 } Str\left( e^{- \pi v(L_{0} + \bar{L}_{0})} \int_{mod (1)} d w \, e^{ 2 \pi i w(L_{0} - \bar{L}_{0})}     \right). \label{usupertr}
\eeq

\vspace{.27 cm}

This latter quantity has an enumerative meaning related to the towers of massive closed string excitations.
 In terms of  effective numbers of closed string states reads
\vspace{.27 cm}

\beq
A_1 =  Vol(\mathcal{D}_1) \lim_{v \rightarrow 0}\, v^{1 - d/2 }  \sum_{ | \Phi \rangle} (-)^{F_{\Phi}} d( \Phi  ) e^{ - \pi v m^{2}_{\Phi} }, \label{u1str}
\eeq

\vspace{.27 cm}

where the sum  is restricted to closed string \emph{physical} states $| \Phi \rangle$, $(L_0 - \bar{L}_0 ) | \Phi \rangle = 0 $.
 Fermionic states are counted with a minus sign, and $d(\Phi)$ is the number of physical polarizations of   $|\Phi \rangle$
of mass $m_{\Phi}$. At a diagrammatical level, the above form suggests  that one can cut the torus handle, and obtain
an equivalent  representation of the genus one vacuum amplitude as a sum of tree level two-points  amplitudes restricted
to \emph{physical} closed string states, (this is illustrated in figure \ref{fig1}).

\vspace{.27 cm}

One-loop vacuum stability, (absence of closed string tachyons), corresponds to the finiteness of the vacuum amplitude
$A_1 $. This implies via eq. (\ref{u1str})  a constraint on the allowed  deviation from  supersymmetry of one-loop stable closed string
vacua. Convergence  of the series of  tree-level two-point functions in (\ref{u1str}) requires an overall Fermi-Bose degeneracy of the string spectrum.
Let us notice the role of the $v > 0$ coordinate, (the imaginary part of the torus modulus $\tau$),
as a ultraviolet cutoff $\Lambda_{uv} = 1/v$  for the mass of the  states  contributing to eq. (\ref{u1str}).
 Equality in   eq. (\ref{u1str}) holds in the ultraviolet limit $v \rightarrow 0$,
however  one can also consider this relation  for small $v > 0$.
It turns out  that  the error term is under control, and it goes to zero \emph{polynomially}
in the $v \rightarrow 0$ limit. A remarkable fact is that the vanishing rate  in the error estimate
is intimately connected to the Riemann hypothesis.
This was discovered for modular functions  $f = f(w,v)$ of rapid decay by Zagier \cite{Za1}:

\vspace{.27 cm}

\begin{figure}
\begin{center}
\includegraphics[width = 10cm]{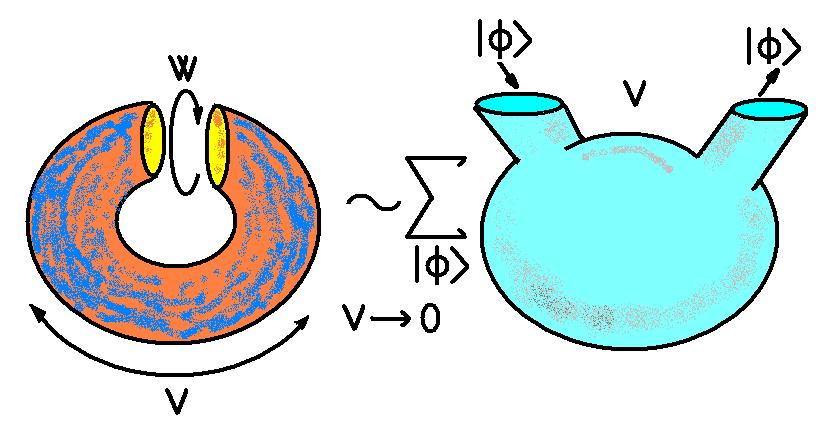}  
\caption{The uniform distribution theorem applied to the one-loop vacuum amplitude (on the left), gives an ultraviolet representation
for this vacuum amplitude in terms of a sum over physical states of tree level (sphere) two-points amplitudes.
Diagrammatically this relation prescribes to cut the torus handle thus creating a sphere with two marking points.
Uniform distribution  theorem instructs  to sum over physical closed string states through the two marked points.} \label{fig1}
\end{center}
\end{figure}

\beq
 \int_{0}^{1} dw f(w,v) \sim \frac{1}{Vol(\mathcal{D}_1)} \int_{\mathcal{D}} dw dv \, v^{-2} \, f(w,v) + O(v^{1 - \frac{\Theta}{2}}) \qquad v \rightarrow 0, \label{uerror}
\eeq

\vspace{.27 cm}

where  $\Theta$ is the superior of the real part of the non trivial zeros of the Riemann zeta function,
($\Theta = \frac{1}{2}$ if and only if the Riemann hypothesis is true, while so far one can prove that $\frac{1}{2} \le  \Theta < 1$).
The asymptotic (\ref{uerror}) implies that  an independent result on the error term
would prove (or disprove) the Riemann hypothesis.
The above relation has been proved to hold also for modular functions of polynomial growth for $v \rightarrow \infty$, appearing in type II
string theory \cite{ACER},\cite{C2}. Eq. (\ref{uerror}) is  expected to hold also from certain  arguments in heterotic strings \cite{ACER},\cite{C2},
although it is  an open  challenge to prove it in this latter case \cite{C2}.

\vspace{.27 cm}

In one-loop vacua,  above considerations lead  to an interesting relation  between ultraviolet
behavior  of  closed string spectra
and the Riemann hypothesis \cite{ACER}. This is given by the following asymptotic:

\vspace{.27 cm}

\beq
\Lambda_{uv}^{ d/2 - 1 }  \sum_{ | \Phi \rangle} (-)^{F_{\Phi}} d( \Phi  ) e^{ - \pi  m^{2}_{\Phi}/\Lambda_{uv}^{2}} \sim \frac{A_1}{Vol(\mathcal{D}_{1})}  + O(\Lambda_{uv}^{\Theta - 1}) \qquad \Lambda_{uv} \rightarrow \infty, \nn
\eeq

which implies that differences between bosonic and fermionic degrees of freedom oscillate
with the frequencies given by  imaginary parts of the non trivial zeros of the Riemann zeta function \cite{ACER}.
Moreover, asymptotic supersymmetry is maximal if and only if the Riemann hypothesis is true \cite{ACER}.

\vspace{.27 cm}

 It is now  worth to explain the Iwasawa decomposition  origin  of
the two coordinates $w$ and $v$ in $\tau = w + iv$.
The upper complex plane $\mathcal{H}_1$ is isomorphic to the Lie coset $\mathcal{H}_{1} \sim Sp(2,\mathbb{R}) / SO(2,\mathbb{R})$.
Given a matrix $\begin{pmatrix} a & b \\ c & d  \end{pmatrix}$ in this coset,
the bijective map is given by

\vspace{.27 cm}

\beq
\tau = (ai + b) (c i + d)^{-1}. \label{map}
\eeq

\vspace{.27 cm}

The Iwasawa decomposition allows to write a symplectic matrix $g \in Sp(2,\mathbb{R})$ as

\beq
g =
\begin{pmatrix} 1 & w \\ 0 & 1  \end{pmatrix}
\begin{pmatrix} v^{1/2} & 0 \\ 0 & v^{-1/2}  \end{pmatrix}
\begin{pmatrix} \cos \vartheta &  \sin \vartheta  \\  - \sin \vartheta & \cos \vartheta  \end{pmatrix}
\qquad w \in \mathbb{R}, v > 0, \vartheta \, {\rm{mod}}(2\pi) \label{Iw1},
\eeq

From the isomorphism $\mathcal{H}_1 \sim Sp(2,\mathbb{R}) / SO(2,\mathbb{R})$
given by the map in (\ref{map}), one thus finds  $\tau = w + i v$.
Therefore, the horocycle flow along the $w$ coordinate
uplifted  in the homogenous space $Sp(2,\mathbb{R})/SO(2,\mathbb{R})$ is generated by
 unipotent elements, given by the upper triangular
matrix of the Iwasawa decomposition (\ref{Iw1}).
On the other hand, $v$  corresponds  to
the coordinate of the  abelian part in the Iwasawa decomposition (\ref{Iw1}).

\vspace{.27 cm}

\section{Genus two: uniformization of unipotent flows viz cutting the  amplitude handle(s)}\label{genus2}

\vspace{.27 cm}

In this section we illustrate in some details the genus $g = 2$ case,
while higher genera are discussed  in the next sections.
The moduli space of genus-two compact Riemann surfaces $\mathcal{M}_2$
is isomorphic to $ Sp(4,\mathbb{Z})\backslash \mathcal{H}_2$. $\mathcal{H}_2$
is the genus two Siegel half space, given by complex symmetric two by two matrices $\tau$, with
 positive definite imaginary part.
$\mathcal{H}_2$ is isomorphic to $Sp(4,\mathbb{R})/ (SO(4,\mathbb{R}) \cap Sp(4,\mathbb{R}))$,
the symplectic matrices over the orthosymplectic  ones.

\vspace{.27 cm}

By Iwasawa decomposition, each element $m$ of the above coset can be written as $m = UA$,
where $U$ is a unipotent matrix and $A$ is a  abelian matrix, (we refer to our works  \cite{CC1},\cite{CC2} for notations and proofs \cite{CC2} used throughout the rest
of this paper).

\vspace{.27 cm}

 One finds for the genus-two period matrix $\tau_{(2)}$

\vspace{.27 cm}

\beq
\tau_{(2)} =
\begin{pmatrix}
w_1 + i(v_1 + u^2 v_2) & w_2 + i u v_2 \\
w_2 + i u v_2  &  w_3 + i v_2
\end{pmatrix}. \label{tau2}
\eeq

\vspace{.27 cm}

\begin{figure}
\begin{center}
\includegraphics[width = 8cm]{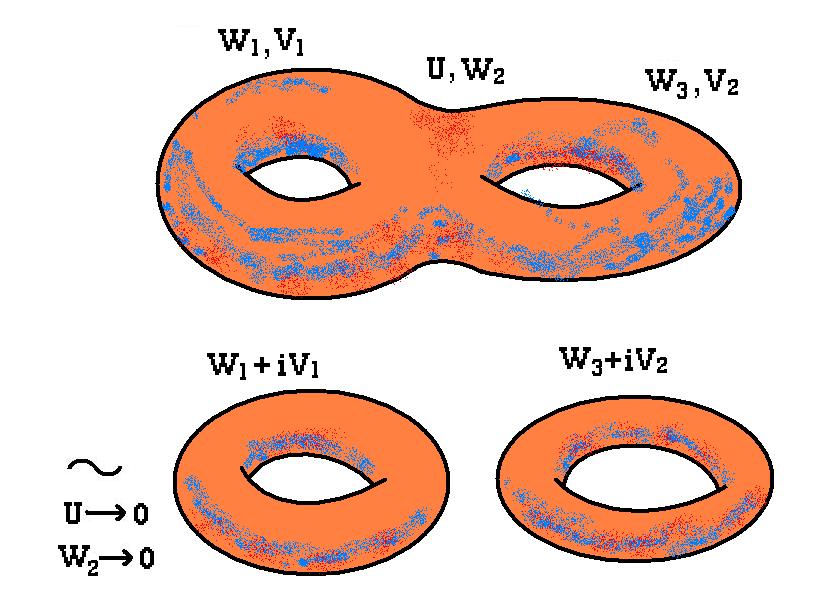}
\caption{In the Iwasawa parametrization of the genus two Siegel half-space $\mathcal{H}_2$, the degenerating limit
is realized by sending to zero off-diagonal unipotent moduli \cite{CC2}.}\label{SFig2}
\end{center}
\end{figure}

Thus  a  genus-two Riemann surface degenerates
into  two genus-one Riemann  surfaces when both the off-diagonal unipotent moduli  $u$ and $w_2$ go to zero,
(figure \ref{SFig2})

\beq
\tau_{(2)} =
\begin{pmatrix}
w_1 + i(v_1 + u^2 v_2) & w_2 + i u v_2 \\
w_2 + i u v_2  &  w_3 + i v_2
\end{pmatrix}
\rightarrow
\begin{pmatrix}
w_1 + iv_1  &   0 \\
   0    &  w_3 + i v_2
\end{pmatrix}
=
\begin{pmatrix}
\tau_1   &   0 \\
   0    &   \tau_3
\end{pmatrix}.    \label{degenerate211}
\eeq

\vspace{.27 cm}

A genus-two  closed string amplitude is given by a  modular integral

\vspace{.27 cm}

\beq
A_2  =  \int_{\mathcal{D}_2} d\mu_2  \, f_{2}(\tau_{(2)}), \nn
\eeq
\vspace{.27 cm}

with $\mathcal{D}_2 \sim  Sp(4,\mathbb{Z})  \backslash \mathcal{H}_2$ is a fundamental region
of the genus-two modular group $\Gamma_2 \sim Sp(4,\mathbb{Z})$.
Uniform distribution theorem, (Theorem 2 in \cite{CC2}), allows to rewrite the genus-two modular integral
as the integral over  the corank-one component of the  $\mathcal{H}_2$ boundary  of
the following  $f_2$   unipotent average

\begin{figure}
\begin{center}
\includegraphics[width = 8cm]{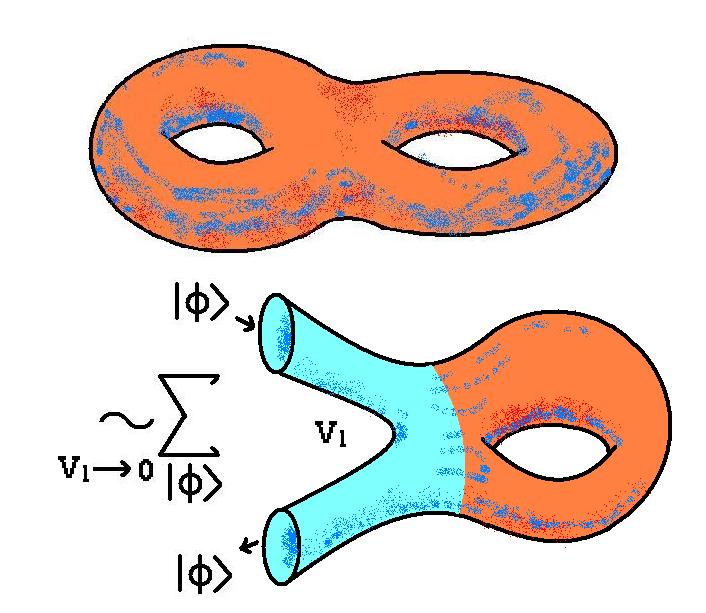}
\caption{Uniformization theorem, (Theorem 2 in \cite{CC2}), allows to cut an handle of a genus-two vacuum amplitude, with the prescription
  of  summing  over physical states flowing through the
two marked points.}\label{SFig3}
\end{center}
\end{figure}

\vspace{.27 cm}

\beq
A_2 =  \frac{Vol(\mathcal{D}_2)}{2 Vol(\mathcal{D}_1)} \lim_{v_1 \rightarrow 0} \int_{\mathcal{D}_1} d\mu_{1} \, \int_{mod(1)}dw_1 \int d w_2 \, d u f_{2}(\tau_{(2)}).  \label{uni21}
\eeq

\vspace{.27 cm}

On the other hand, Theorem 1 in \cite{CC2}, ensures  the following  unipotent average function, defined on $\mathcal{H}_1 \times \mathbb{R}_{> 0}$

\beq
f_{1}(\tau_3, v_1 ) = \int_{mod(1)}dw_1 \int d w_2 \, d u f_{2}, \label{f1}
\eeq

to be  invariant under  $SL(2,\mathbb{Z})$ modular transformations
on $\tau_3$

\beq
\tau_{(2)} =
\begin{pmatrix}
\tau_1 & \tau_2 \\
\tau_2  &   \tau_3
\end{pmatrix},
\qquad  \tau_1 \in \mathcal{H}_1,  \, \, \tau_3 \in \mathcal{H}_{1}. \nn
\eeq

\vspace{.27 cm}

\begin{figure} 
\begin{center}
\includegraphics[width = 8cm]{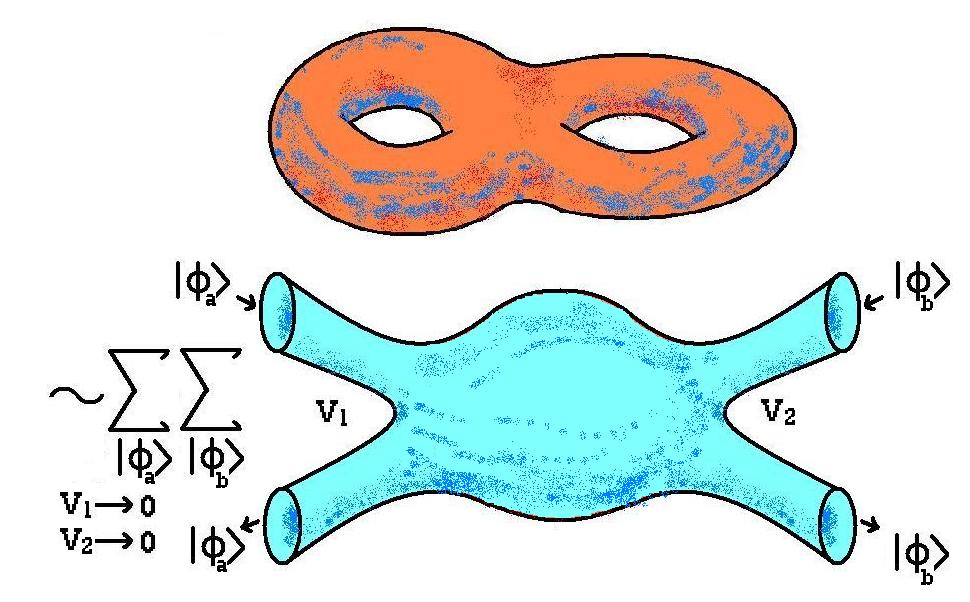}
\caption{By applying twice the uniform distribution theorem for unipotent flows, one can write the genus two vacuum amplitude as
a sum of three level four points amplitudes. This implies that on a stable vacuum,  two-loop infrared finiteness
is translated into a  constraint on the  asymptotic ultraviolet behavior of tree-level four-points  closed string amplitudes.}\label{SFig4}
\end{center}
\end{figure}

From the degenerate limit (\ref{degenerate211}), and the discussion on the  genus-one case  in \S \ref{genus1},
 one sees that integration mod(1) along the coordinate $w_1$ together with the limit $v_1 \rightarrow 0$ in (\ref{uni21}),
diagrammatically correspond to the situation displayed in  figure \ref{SFig3}. This unipotent flow representation corresponds to
cutting one handle of the genus-two vacuum amplitude and replacing it by two external legs, (figure \ref{SFig3}).
This operation  generates a genus-one amplitude with two external legs.
Uniformization theorem \cite{CC2} gives prescription of summing
over all physical external    states $| \Phi \rangle$.
Therefore, infrared finiteness at genus-two order, $|A_2| < \infty$, through eq. (\ref{uni21})
corresponds to finiteness of the one-loop correction to the asymptotic Fermi-Bose degeneracy condition,
described in \S \ref{genus1}. Thus, we have shown that ergodicity results of  unipotent
 flows \cite{CC2} lead to the one-loop correction to the
 \emph{asymptotic supersymmetry} constraint, obtained in \cite{KS}.

 \vspace{.27 cm}

 However, there is another incarnation of the genus-two vacuum amplitude, in terms
 of a sum of \emph{tree-level four-point functions}, (figure \ref{SFig4}).
 This is obtained by applying the uniformization theorem for unipotent flows on
 the $SL(2,\mathbb{Z})$-invariant  function $f_1$ in eq. (\ref{f1}):

 \vspace{.27 cm}

 \beq
A_2 =  \frac{Vol(\mathcal{D}_2)}{2} \lim_{v_1 \rightarrow 0} \lim_{v_2 \rightarrow 0} \int_{mod(1)} d w_3 \, \int_{mod(1)}dw_1 \int d w_2  d u f_{2}(\tau_{(2)}).  \label{uni20}
\eeq

 \vspace{.27 cm}

Diagrammatically, eq. (\ref{uni20}) corresponds to transmute the original genu-two vacuum amplitude in a double sum
of tree level four-point functions,  extended to physical closed string states, (figure \ref{SFig4}).
Level matching for external states flowing  in the external legs follow by integration
mod(1), along the diagonal  unipotent  coordinates $w_1$ and $w_3$ in the period matrix  (\ref{tau2}).
On the other hand,  abelian coordinates $v_1$ and $v_2$
act as ultraviolet cutoffs for the masses of the  external states.

\vspace{.27 cm}

 One can also write an asymptotic expression for $A_2$ when $v_1$ and $v_2$ are both small.
 As in the genus one  case described in \S \ref{genus1},  the error term turns out to be  remarkably connected to the
 Riemann hypothesis \cite{CC1},\cite{CC2}:

 \vspace{.27 cm}

\beq
A_2  \, {\sim}_{\substack{v_1 \rightarrow 0\\v_2 \rightarrow 0}} \, \frac{Vol(\mathcal{D}_2)}{2}  \int_{mod(1)} d w_3 \, \int_{mod(1)}dw_1 \int d w_2  d u f_{2}(\tau_{(2)})
+ O(v_{1}^{2 - \frac{\Theta}{2}}) + O(v_{2}^{1 - \frac{\Theta}{2}}). \nn
\eeq

$\Theta$ is the superior of the real part of the non trivial zeros of the Riemann zeta function,
($\Theta = 1/2$ if and only if the Riemann hypothesis is true).

\vspace{.27 cm}

\section{Uniformizations for closed amplitudes at genus $g= 1,2,3$, and moduli of punctured Riemann surfaces.}
\vspace{.27 cm}

Uniformization results \cite{CC2} concern integrals of automorphic forms on the moduli space of genus $g$ principally polarized abelian varieties (ppav)
$\mathcal{A}_g$. Every point in $\mathcal{A}_g$ describes a $g$-dimensional torus which can
be embedded in a projective space (abelian variety), with principal polarization.
The moduli space of genus $g$ compact Riemann surfaces $\mathcal{M}_g$
is isomorphic to $\mathcal{A}_g$ for $g = 1,2,3$, while for genus $g \ge 4$, $\mathcal{M}_g$ is a subvariety  fully contained in
$\mathcal{A}_g$ of (complex) codimension ${\rm{dim}}(\mathcal{A}_g) - {\rm{dim}}(\mathcal{M}_g)  = \frac{1}{2}(g - 2)(g - 3)$.

\vspace{.27 cm}

Uniformization results for modular integrals of automorphic forms on $\mathcal{A}_g$ \cite{CC2}, when applied to
closed string vacuum amplitudes  connect a genus $g$ vacuum amplitude to sums of  lower genera scattering  amplitudes.
Diagrammatically, one transmutes
 a handle in the Riemann surface into   a pair of marked points.
   For each  cut   handle, the amplitude genus $g$ is lowered  by a unit, while
 two external legs are added  to  the amplitude.  For the moduli space of genus $g$ compact  Riemann surfaces with $n$
marked points $\mathcal{M}_{g, n}$, one has ${\rm{dim}}(\mathcal{M}_{0,n}) = n - 3$ (Riemann sphere with $n$ marked points),
${\rm{dim}}(\mathcal{M}_{1,n}) = n$ (torus with $n$ marked points), ${\rm{dim}}(\mathcal{M}_{g,n}) = 3g - 3 + n$, $g \ge 2$.

\vspace{.27 cm}

For the genus one-amplitude (torus) discussed in \S \ref{genus1}, the uniform distribution of long horocycles  theorem reduces
this amplitude in a sum of sphere amplitudes with two marked points, (figure \ref{fig1}). In the  original torus modulus $\tau = w + iv$,
 integration mod(1) of the unipotent Iwasawa coordinate $w$  forces level matching for  the external closed string states flowing through the two marked points.
The abelian Iwasawa coordinate  $v$ provides  a UV cutoff $\Lambda = 1/v$ for their masses. There are no extra moduli besides $w$ and $v$,
 consistently  with the absence of moduli for Riemann spheres with two marked points, ${\rm{dim}}(\mathcal{M}_{0,2}) = 0$.

 \vspace{.27 cm}

For the genus-two vacuum amplitude discussed in \S \ref{genus2}, uniformization results \cite{CC2} give two representations.
One  is given as a sum of genus-one amplitudes with two marked points, (figure \ref{SFig3}), while the other one
  is given  in terms of  a sum of genus-zero amplitudes with four marked points, (figure \ref{SFig4}).
 The first case is displayed in eq. (\ref{uni21}), with  integrated moduli  $w_1$, $w_2$, $u$ and $w_3 + i v_2$ in the
  periods matrix  $\tau_{(2)}$,  eq. (\ref{tau2}).  Integration on $w_1$ mod(1) ensures level matching for the
 closed string states flowing through the two marked points,  while $w_3 + i v_2$ is the modulus  related to
 the left over handle.
 $w_3 + i v_2$   together with  $w_2$ and $u$ consistently account for  the number of moduli
of tori with two marked points,   ${\rm{dim}}(\mathcal{M}_{1,2}) = 2$, (complex dimension).
For the second representation of $A_2$, displayed in eq. (\ref{uni20}), the genus-two vacuum amplitude is given by
a sum over genus-zero amplitudes with four marked points. The two diagonal unipotent moduli  $w_1$ and $w_3$
integrated mod(1) ensure level matching for states flowing through the two pairs of marked points, (figure \ref{SFig4}).
On the other hand, the off-diagonal unipotent   moduli $w_2$ and $u$ consistently account for the dimension of the moduli space
of spheres with four marked points ${\rm{dim}}(\mathcal{M}_{0,4}) = 1$.

\vspace{.27 cm}

At genus three, the periods matrix in Iwasawa parametrization of $\mathcal{H}_3$ is given by

\vspace{.27 cm}

\bea
\tau_{(3)} &=&
\begin{pmatrix}
w_{11} + i(v_{1} + v_{2}u_{12}^{2} + v_{3}u_{13}^{2}) & \,  w_{12} + i(v_{2}u_{12} + v_{3}u_{12}u_{13}) & \, w_{13} + iv_{3}u_{13} \\
                                      *                & \,  w_{22} + i(v_{2} + v_{3}u_{23}^{2}) & \, w_{23} + iv_{3} u_{23}   \\
*                                                      &              *                       &\,  w_{33} + i v_{3}
\end{pmatrix} \nn \\
&=&
\begin{pmatrix}
\tau_{11} & \tau_{12} \\
\tau_{12}^{t} & \tau_{22}
\end{pmatrix},
\qquad \tau_{11} \in \mathcal{H}_{1}, \tau_{22} \in \mathcal{H}_{2}. \nn
\eea

\begin{figure} 
\begin{center}
\includegraphics[width = 14cm]{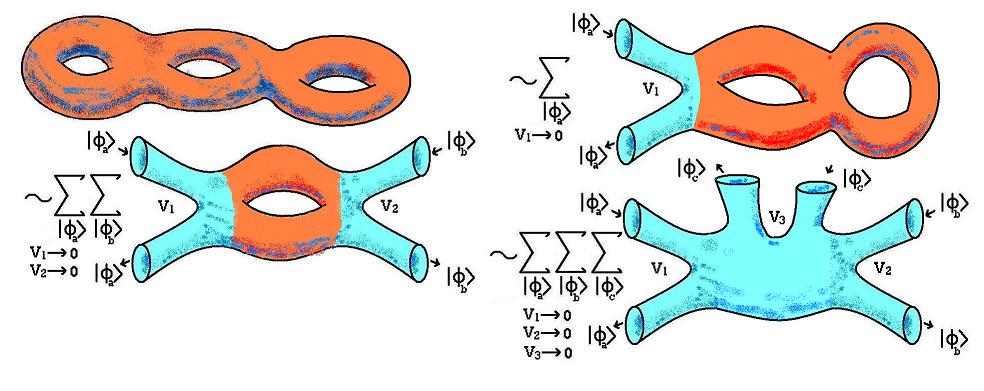}  
\caption{Three different  representations for  a genus three closed string  vacuum amplitude, that follow from the uniformization theorems of unipotent flows
\cite{CC2}.}\label{SFig5}
\end{center}
\end{figure}

\vspace{.27 cm}

where $*$ entries are given by symmetry. Uniformization theorems  in  \cite{CC2} give three alternative representations for the
 closed string vacuum amplitude $A_3$, (figure \ref{SFig5})

\vspace{.27 cm}

By cutting one handle, one can write $A_3$ as a sum of genus-two amplitudes with two marked points,
while by cutting two handles $A_3$ is given by a sum of genus-one amplitudes with four marked points.
 Finally, by cutting three handles $A_3$ can be expressed as  a sum of genus-zero amplitudes with six marked points.

\vspace{.27 cm}

The representation in terms of genus-two amplitudes is given by

\beq
A_{3} \,  {\sim}_{\substack{v_1 \rightarrow 0}} \, \frac{Vol(\mathcal{D}_{3})}{2Vol(\mathcal{D}_{2})}\int_{\mathcal{D}_2} d\mu_{2} \int_{mod(1)}dw_{11} \int dw_{12} dw_{13} du_{12} du_{13} f_{3}(\tau_{(3)}). \nn
\eeq

 The diagonal unipotent modulus  $w_{11}$  integrated mod(1) enforces
level matching for the states flowing thorough the two marked points. The off-diagonal unipotent
moduli $w_{12}, w_{13}$, $u_{12}, u_{13}$,
together with  $\tau_{22} \in  \mathcal{H}_2$, (the periods matrix  of the two-leftover handles),
 account for the dimension of the moduli space of genus-two Riemann surfaces
with two marked points ${\rm{dim}}(\mathcal{M}_{2,2}) = 5$.

\vspace{.27 cm}

The representation  of $A_3$ as  a sum of genus-one amplitudes  with four marked points is given by

\vspace{.27 cm}

\bea
A_{3}  {\sim}_{\substack{v_1 \rightarrow 0\\v_2 \rightarrow 0}} \frac{Vol(\mathcal{D}_{3})}{4Vol(\mathcal{D}_{1})} \int_{\mathcal{D}_1} d\mu_{1}\int_{mod(1)}dw_{11} \int_{mod(1)}dw_{22}\int dw_{12} dw_{13} dw_{23} du_{12} du_{13} du_{23} f_{3}(\tau_{(3)}), \nn
\eea

\vspace{.27 cm}

Diagonal unipotent moduli  $w_{11}$ and $w_{22}$ integrated mod(1) select physical states through the two pairs
of marked points, (figure \ref{SFig6}). Off-diagonal unipotent coordinates $w_{12},w_{13},w_{23}$, $u_{12},u_{13},u_{23}$
and  $\tau_{(1)} \in \mathcal{H}_1$ correctly  account  for the dimension of the moduli space of tori with four
marked points,  ${\rm{dim}}(\mathcal{M}_{1,4}) = 4$.

\vspace{.27 cm}

The representation of  $A_3$  as a  sum  of genus-zero  amplitudes with six marked points is given by

\vspace{.27 cm}

\bea
A_{3}  {\sim}_{\substack{v_1 \rightarrow 0\\v_2 \rightarrow 0\\ v_3 \rightarrow 0}} \frac{Vol(\mathcal{D}_{3})}{8} \int_{mod(1)}dw_{11} \int_{mod(1)}dw_{22} \int_{mod(1)}dw_{33}   \int dw_{12} dw_{13} dw_{23} du_{12} du_{13} du_{23} f_{3}(\tau_{(3)}). \nn
\eea

\vspace{.27 cm}

Diagonal unipotent moduli $w_{11}$, $w_{22}$, $w_{33}$ integrated mod(1) select physical closed string states through
the three pairs of marked points (figure \ref{SFig5}). Off-diagonal unipotent moduli $w_{12},w_{13}, w_{23}$, $u_{12},u_{13}, u_{23}$
 consistently account  for the dimension of the moduli space of spheres with six marked points, ${\rm{dim}}(\mathcal{M}_{0,6}) = 3$.

\vspace{.27 cm}
\section{Uniformization at every genera: closed string hints for relevance of unipotent flows for the Schottky problem.}
\vspace{.27 cm}

Uniformization results \cite{CC2} concern  integrals of $Sp(2g,\mathbb{Z})$ automorphic forms on the
moduli space of genus $g$ principally polarized abelian varieties $\mathcal{A}_g$.
 The  moduli space of genus $g$ compact Riemann surfaces $\mathcal{M}_g$
is isomorphic to $\mathcal{A}_g$ for $g = 1,2,3$.
 For genus $g \ge 4$, $\mathcal{M}_g$ is a subvariety  fully contained in
$\mathcal{A}_g$, of (complex) codimension ${\rm{dim}}(\mathcal{A}_g) - {\rm{dim}}(\mathcal{M}_g)  = \frac{1}{2}(g - 2)(g - 3)$.
The embedding of $\mathcal{M}_g$ in $\mathcal{A}_g$ is called the Schottky locus $\mathcal{S}_g$,
and the problem of  its complete  characterization for every genus is still  wide open, (see for example \cite{Gr}).
Genus four is the first non-trivial case, and the Schottky locus $\mathcal{S}_4$ is a divisor in
$\mathcal{A}_4$, (it is of complex codimension one). In this case,  $\mathcal{S}_4$  is fully characterized by
the vanishing of a $16$-degree polynomial in the theta nulls, (the Igusa form $I_{4}$).

\vspace{.27 cm}

\begin{figure} 
\begin{center}
\includegraphics[width = 16cm]{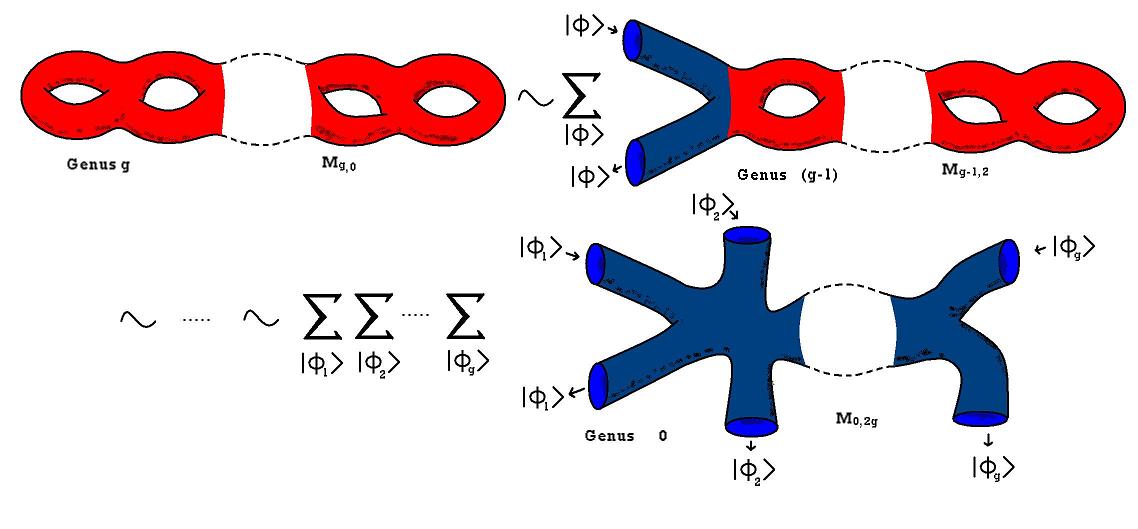}  
\caption{Uniformization relations for  a vacuum amplitude of arbitrary genus.}\label{SFig6}
\end{center}
\end{figure}

From a string theory diagrammatic  perspective, it seems that nothing special occurs at  genus $g = 4$, or higher.
 This leads us to conjecture for the  cutting-handles  procedure to hold at every
genera

\beq
\mathcal{A}_{g,0} = \sum_{|\Phi\rangle}\mathcal{A}_{g-1, 2} =\dots =  \sum_{a=1}^{g} \, \sum_{|\Phi_{a}\rangle}\mathcal{A}_{0, 2g}( |\Phi_{1}\rangle, \dots |\Phi_{g}\rangle), \nn
\eeq

where

\beq
\mathcal{A}_{g,2n}   = \int_{\mathcal{M}_{g, 2n}} d\mu_g  \,  f_{g,n}( |\Phi_{1}\rangle, \dots |\Phi_{n}\rangle ), \nn
\eeq

is the $2n$-points genus $g$ closed string amplitude with external states $|\Phi_{1}\rangle, \dots |\Phi_{n}\rangle$ arranged as in figure \ref{SFig6}.

\vspace{.27 cm}

 This suggests the interesting possibility that results in the dynamics
of unipotent flows may be used to treating    modular integrals
restricted to the Schottky locus $\mathcal{S}_g$. This
is of interest for the still open problem of constructing  closed string amplitudes of arbitrary genus.
Moreover, one may use unipotent flows to connect recently proposed genus three \cite{CDPvG}, (and  genus higher then  three \cite{CDPvG}, \cite{Gr2},  \cite{GSM1},\cite{GSM2},\cite{MV1}, \cite{MV2}, \cite{MV3}),
closed string amplitudes to the  genus-two expressions given in  \cite{DHP}.
All the above  possibilities are presently under  investigation \cite{CC3}, \cite{CCDP}.

\section{Summary and Future Directions}

In this paper we have applied to string theory  mathematical results
on uniform distribution of unipotent flows
we have recently  proved \cite{CC2}.
We  obtained  conditions of perturbative stability \emph{at all genera} in non-supersymmetric
closed string vacua,  in terms of  \emph{solely}  closed string  three-level diagrams.
 The key is provided by ergodic properties of unipotent flows,
which allow to study closed strings asymptotic  UV properties,
by probing boundary  components of $\mathcal{M}_g$, (the genus $g$ moduli space
of compact Riemann surfaces).
Diagrammatically,  unipotent flows asymptotics translate
into prescriptions of cutting  amplitudes  handles,
while  summing over physical states flowing through
pairs of marked points. Remarkably,  those asymptotics have error
 estimates related to  the Riemann hypothesis \cite{CC1},\cite{CC2}.

\vspace{.27 cm}

Our analysis extends to \emph{all genera} in perturbation theory, and generalizes  previous
 results at genus-one level \cite{KS}, (see also \cite{Di}).
It would be interesting to check whether our condition  on ultraviolet perturbative
stability of a closed string vacuum  make contact with stability  conditions  in closed string field theory.
Moreover, our closed string theory constraints may be  of interest for  higher spins theories, (see for example  \cite{FS},\cite{ST}).
We also notice some formal similarities between our closed string
cutting procedures and analytic cuts procedures in quantum field theory \cite{Dix}.
It would be interesting to check whether those  analogies point to some deep relations.
We also would like to mention that our cutting techniques seem to suit for
connecting loops vacuum closed string  amplitudes to the most  general closed string  tree-level amplitudes.
This may be achieved by  taking appropriate limits for vertex operators on the punctured Riemann surface \cite{Sa}. Progress in this direction,
would provide explicit expressions in  type II A, type II B
and Type 0 closed string theories for genus two and genus three closed string amplitudes \cite{Sa}.

\vspace{.27 cm}

 From the mathematical side, our diagrammatic prescriptions on closed string amplitudes
 suggest a relevance of unipotent flows for the Schottky problem, and for the problem
 of defining the superstring amplitude at arbitrary genera.
 These issues are  presently under investigation \cite{CC3}, \cite{CCDP},
 in two distinct directions. On the one hand, we would like to
 obtain uniformization  results for integrals of automorphic functions
 over the Schottky locus.  Existence of such an interesting  possibility is suggested  diagrammatically
  by string theory \cite{CC3}.
 On the other hand, by using  uniformization result  in \cite{CC2},
 we would like to be able  to connect recently  proposed genus $g = 3,4,5$ closed string amplitudes
 \cite{CDPvG},\cite{DPvG},\cite{DPGC}, \cite{Gr},\cite{GSM1},\cite{GSM2},\cite{Mo},\cite{MV1},\cite{MV2},\cite{MV2}   to the
 genus two superstring  amplitudes given in  \cite{DHP}. This would provide  consistency checks for the
 proposed genus $g \ge 3$ superstring  amplitudes \cite{CCDP}.

\vspace{ .27 cm}

\section*{Acknowledgments}
During  elaboration of the ideas and results involved in  this work,  MC was supported at various stages by:
 Superstring Marie Curie Training
Network under the contract MRTN-CT-2004-512194
at the Hebrew University of Jerusalem,  Visiting fellowship at the ESI
Schroedinger Center for Mathematical Physics in Vienna,
Italian MIUR-PRIN contract 20075ATT78 at the University
of Milano Bicocca, Visiting fellowship of  the  Theory Unit at CERN, and by a
"Angelo Della Riccia" fellowship at the University of Amsterdam.


\begin{thebibliography}{abc}







\bibitem[ACER]{ACER}
  C.~Angelantonj, M.~Cardella, S.~Elitzur and E.~Rabinovici,
  ``Vacuum stability, string density of states and the Riemann zeta function,''
  JHEP {\bf 1102}, 024 (2011)
  [arXiv:1012.5091 [hep-th]].

\bibitem[C1]{C1}  M.~Cardella,
  ``A novel method for computing torus amplitudes for $\mathbb{Z}_{N}$
  orbifolds without the unfolding technique,''
  JHEP {\bf 0905} (2009) 010.


\bibitem[C2]{C2}
M.~A.~Cardella,
  ``Error Estimates in Horocycle Averages Asymptotics: Challenges from String
  Theory,''
  arXiv:1012.2754 [math.NT].



\bibitem[CC1]{CC1}   S. Cacciatori  M.~Cardella,
  ``Equidistribution rates,  closed string amplitudes, and the Riemann
  hypothesis,''
  JHEP12 (2010) 025.

\bibitem[CC2]{CC2}
  S.~L.~Cacciatori and M.~A.~Cardella,
  ``Uniformization, Unipotent Flows and the Riemann Hypothesis,''
  arXiv:1102.1201 [math.NT].




\bibitem[CC3]{CC3}
  S.~L.~Cacciatori and M.~A.~Cardella,
  ``Unipotent Flows reductions of modular integrals over the Schottky locus,''
  In progress.


\bibitem[CCDP]{CCDP}
  S.~L.~Cacciatori and M.~A.~Cardella, F. Dalla Piazza,
  ``Genus-two reductions of genus-three superstring amplitudes via unipotent flows,''
  In progress.





\bibitem[CDPvG]{CDPvG}
 S. L. Cacciatori and F. Dalla Piazza, ``Two loop superstring amplitudes and S6
representations,'' Lett. Math. Phys. 83, 127 (2008); \\
 S. L. Cacciatori, F. Dalla Piazza and B. van Geemen, ``Modular Forms and Three Loop
Superstring Amplitudes,'' Nucl. Phys. B 800, 565 (2008); \\
S. L. Cacciatori, F. D. Piazza and B. van Geemen, ``Genus four superstring measures,'' Lett.
Math. Phys. 85, 185 (2008).


\bibitem[Di]{Di}
  K.~R.~Dienes,
  ``Modular invariance, finiteness, and misaligned supersymmetry: New
  constraints on the numbers of physical string states,''
  Nucl.\ Phys.\  B {\bf 429}, 533 (1994)

\bibitem[Dix]{Dix}
  L.~J.~Dixon,
  ``Calculating scattering amplitudes efficiently,''
  arXiv:hep-ph/9601359.



\bibitem[DPvG]{DPvG}
F. D. Piazza and B. van Geemen, ``Siegel modular forms and finite symplectic groups,'' Adv.
Theor. Math. Phys. 13 (2009) 1771-1814.




\bibitem[DHP]{DHP}
E. D'Hoker, D.H. Phong, ``Two-Loop Superstrings I, Main Formulas'', Phys. Lett. B 529
(2002) 241-255; \\
 E. D'Hoker, D.H. Phong, ``Two Loop Superstrings II. The Chiral measure on Moduli Space'',
Nucl. Phys. B636 (2002) 3-60; \\
 E. D'Hoker, D.H. Phong, ``Two Loop Superstrings III. Slice Independence and Absence of
Ambiguities'', Nucl. Phys. B636 (2002) 61-79; \\
 E. D'Hoker, D.H. Phong, ``Two-Loop Superstrings IV: The Cosmological Constant and
Modular Forms'', Nucl. Phys. B 639 (2002) 129-181; \\
 E. D'Hoker, D.H. Phong, ``Asyzygies, Modular Forms, and the Superstring Measure I'', Nucl.
Phys. B 710 (2005) 58-82; \\
 E. D'Hoker, D.H. Phong, ``Asyzygies, Modular Forms, and the Superstring Measure II'',
Nucl. Phys. B 710 (2005) 83-116.


\bibitem[DPGC]{DPGC}
  F.~D.~Piazza, D.~Girola and S.~L.~Cacciatori,
  ``Classical theta constants vs. lattice theta series, and super string
  partition functions,''
  JHEP {\bf 1011}, 082 (2010).


\bibitem[DS]{DS} S. G. Dani and J. Smillie,
``Uniform distribution of horocycle orbits for Fuchsian groups,''
 Duke Math. J. 51 (1984), 185-194.

\bibitem[EW]{EW}  Manfred Einsiedler and Thomas Ward,
``Ergodic Theory with a view towards Number Theory,''
Springer Graduate Text in Mathematics Volume 259.

\bibitem[EMV]{EMV} M. Einsiedler, G. Margulis, A. Venkatesh,
``Effective equidistribution for closed orbits of semisimple groups on homogeneous spaces,''
Inventiones Mathematicae, Volume 177, Number 1, 137-212,.


\bibitem[ELMV]{ELMV}
Manfred Einsiedler, Elon Lindenstrauss, Philippe Michel, Akshay Ven,
``Distribution of periodic torus orbits and Duke's theorem for cubic fields,''
to appear in Annals of Mathematics;
[arXiv:0708.1113 [math.DS]].


\bibitem[Fu]{Fu} H. Furstenberg, ``The Unique Ergodicity of the Horocycle Flow, Recent Advances in Topological Dynamics,'' A. Beck (ed.), Springer Verlag Lecture Notes, 318 (1972), 95-115.


\bibitem[FS]{FS}
  D.~Francia and A.~Sagnotti,
  ``Free geometric equations for higher spins,''
  Phys.\ Lett.\  B {\bf 543}, 303 (2002)
  [arXiv:hep-th/0207002].


\bibitem[Gr]{Gr} S. Grushevsky,
``The Schottky Problem,''
To appear in the proceedings of "Classical algebraic geometry today" workshop (MSRI, January 2009)


\bibitem[Gr2]{Gr2} S. Grushevsky,
 ``Superstring scattering amplitudes in higher genus,'' Commun. Math. Phys.
287, 749 (2009).




\bibitem[GSM1]{GSM1}
  S.~Grushevsky and R.~S.~Manni,
  ``The superstring cosmological constant and the Schottky form in genus 5,''
  arXiv:0809.1391 [math.AG].

\bibitem[GSM2]{GSM2}
  S.~Grushevsky and R.~Salvati Manni,
  ``The vanishing of two-point functions for three-loop superstring scattering
  amplitudes,''
  Commun.\ Math.\ Phys.\  {\bf 294}, 343 (2010).


\bibitem[He]{He}
 Gustav A. Hedlund,
``Fuchsian groups and transitive horocycles''
Duke Math. J. Volume 2, Number 3 (1936), 530-542.





\bibitem[KS]{KS}
  D.~Kutasov and N.~Seiberg,
``Number Of Degrees Of Freedom, Density Of States And Tachyons In String
  Theory And Cft,''
  Nucl.\ Phys.\  B {\bf 358}, 600 (1991).

\bibitem[Mo]{Mo}
 A. Morozov, ``NSR Superstring Measures Revisited,''
  JHEP {\bf 0805} (2008) 086.


\bibitem[MV1]{MV1}
 M.~Matone and R.~Volpato,
  ``Getting superstring amplitudes by degenerating Riemann surfaces,''
  Nucl.\ Phys.\  B {\bf 839}, 21 (2010).


\bibitem[MV2]{MV2}
  M.~Matone and R.~Volpato,
  ``Superstring measure and non-renormalization of the three-point amplitude,''
  Nucl.\ Phys.\  B {\bf 806}, 735 (2009).


\bibitem[MV3]{MV3}
  M.~Matone and R.~Volpato,
  ``Higher genus superstring amplitudes from the geometry of moduli spaces,''
  Nucl.\ Phys.\  B {\bf 732}, 321 (2006).



\bibitem[Ra]{Ra} M. Ratner,  ``Distribution rigidity for unipotent actions on homogeneous spaces,''  Bull. Amer. Math. Soc. (N.S.) Volume 24, Number 2 (1991), 321-325; \\ M. Ratner,  ``Raghunathan's topological conjecture and distributions of unipotent flows,''  Duke Math. J. Volume 63, Number 1 (1991), 235-280.



\bibitem[Sa]{Sa} A. Sagnotti, Private Communication.



\bibitem[ST]{ST}
  A.~Sagnotti and M.~Tsulaia,
  ``On higher spins and the tensionless limit of string theory,''
  Nucl.\ Phys.\  B {\bf 682}, 83 (2004)
  [arXiv:hep-th/0311257].



\bibitem[Za1]{Za1}
D. Zagier, ``Eisenstein Series and the Riemann zeta function, Automorphic Forms, Representation Theory and Arithmetic,'' Studies in Math. Vol. 10, T.I.F.R., Bombay, 1981, pp. 275-301.


\bibitem[Za2]{Za2} D. Zagier,  ``The Rankin-Selberg method for authomorphic functions which are not of rapid decay,''
in J. Fac. Sci. Tokyo 1981.


\end{thebibliography}
\end{document}